\documentclass[aps,prl,
showpacs]{revtex4}
\usepackage{amsmath}
\usepackage{amssymb}
\usepackage{graphicx}
\usepackage{bm}

\begin{document}

\title{CASIMIR-LIFSHITZ FORCES AND ENTROPY}
\author{L.~P.~PITAEVSKII}
%\date{\today}
\address{INO-CNR BEC Center, Department of Physics, University of Trento, I-38123 Povo,
Trento, Italy;\\
Kapitza Institute for Physical Problems, Kosygina 2, 119334 Moscow, Russia.\\
E-mail:lev@science.unitn.it}

\begin{abstract}
It is shown that the violation of the positiveness of the entropy  
% Nernst's theorem  for ordered bodies 
due to the Casimir-Lifshitz interaction claimed in several papers is an artifact related to  an improper interpretation  of the ''Casimir entropy'', which actually is a difference of two positive terms.
It is explained that at definite condition this ''Casimir entropy''  must be negative. A direct derivation of the low temperature behavior of the surface entropy of a  metallic surface  in conditions of the anomalous skin effect is given and singular temperature dependency of this quantity is discussed. In conclusion a hydrodynamic  example of the entropy of a liquid film is considered. It occurs that the entropy of a film of finite thickness and a liquid half-space behave differently  at $T \to 0$. 
\end{abstract}

\maketitle
%\bodymatter
\section{Physical meaning and properties of the ''Casimir entropy''\label{A}}
Recently much attention has been devoted to the calculation of the entropy related to the Casimir-Lifshitz forces\cite{Sernelius05,Bezzera02,Hoye03,Bostrom04,Bezzera04,Svetovoy05,Mostepanenko06,Svetovoy06}. The common point of these papers is the calculation of the so-called ''Casimir entropy'', which is defined as
\begin{equation}
S_{C}(T,l)=-\int_{l}^{\infty}\left (\frac{\partial f}{\partial T}\right )_{l}dl\; ,
\label{SC}
\end{equation}
where $f(T,l)$ is the force between bodies and  $l$ is the distance between bodies. (I am using units such that $k_{B}=\hbar=1$. In my notation $f$ is negative for attraction.) It was taken for granted that $S_{C}$ is the proper definition of  the entropy and must possess all properties of the entropy, particularly it must be positive.  I will present here considerations that it is not so in general.

It is important to recognize that the Lifshitz theory only gives rigorously  the quantity for which it was developed, i. e. the force $f(T,l)$ for $l \gg d$,  where $d$ is the interatomic distance\cite{Lif,DLP61}. It is impossible to calculate in this theory  full free energy or entropy in terms of dielectric properties of bodies, because these quantities are mainly defined  by short-distances interactions and are divergent in the Lifshitz approximation.  According to thermodynamics
\begin{equation}
f=-\left (\frac{\partial F}{\partial l}\right )_{T}\; ,
\label{f}
\end{equation}
where $F$ is the free energy.
Differentiation with respect to $T$ gives
\begin{equation}
\left (\frac{\partial S}{\partial l}\right )_{T}=\left ( \frac{\partial f}{\partial T}\right)_{l}\; .
\label{S}
\end{equation}
The derivative $(\partial S/\partial l)_{T}$ is the {\it only} information about entropy  given by the Lifshitz theory. For {\it regular} bodies (quantum fluids and their mixtures, regular solids) Nernst's theorem demands that $S(T,l) \to 0$ as $T \to 0$. Thus, in this limit
\begin{equation}
\frac{\partial S}{\partial l}=\frac{\partial f}{\partial T} \to 0\; .
\label{N}
\end{equation}
 Notice that the limit $T \to 0$ must be taken at constant $l$. 
Integration gives
\begin{equation}
\Delta S(T,l,l_0)=S(T,l)-S(T,l_0)=\int^{l}_{l_0}\frac{\partial f}{\partial T}dl\; .
\label{DS}
\end{equation}
Of course, again $\Delta S \to 0$, when $T \to 0$. This is the only consequence of the theorem we can derive. To my knowledge this condition and condition (\ref{N}) were not violated in any {\it proper} calculations for regular bodies, i. e. in calculations where  {\it meaningful} models were solved {\it correctly}. However, $\Delta S$ is a {\it difference} of two positive terms and it is impossible to say anything about its sign. There are also no restrictions on the sign of the specific heat difference which can be obtained from (\ref{DS}):
\begin{equation}
T\frac{\partial \Delta S}{\partial T}=C_V(T,l)-C_V(T,l_0)=T\int^{l}_{l_0}\frac{\partial^2 f}{\partial T^2}dl\; .
\label{DC}
\end{equation}

The ''Casimir entropy'' (\ref{SC}) can be obtained from (\ref{DS}) choosing $l_{0}=\infty$:
\begin{equation}
S_{C}(T,l)=\int^{l}_{\infty}\frac{\partial f}{\partial T}dl=S(T,l)-S(T,l=\infty)\; .
\label{SC1}
\end{equation}

In general, it is impossible to say anything about the sign of $S_C$. However, sometimes 
 it is possible to connect the appearance  of a negative $S_C$ with the temperature dependence of $S_C$ and $\partial S_C/\partial l$ as $T \to 0$. The point is that the temperature dependence of $\partial S_C/\partial l$ is the same as the first term in $S_C$. However, the situation with the second term is different. The limit $l \to \infty $ results in the appearance  of two free surfaces of the bodies. They are new physical objects and their thermodynamic properties
 can be different from the properties of a finite gap between bodies. If it happens in the limit $T \to 0$ that $|S_C| \gg |\partial S_C/\partial l|$, then  the second term in (\ref{SC1}) dominates and $S_C<0$. In this case $S_C$ must obviously be $l$-independent. On the contrary, the leading at $T \to 0$ $l$-dependent term  is originated from the first term and must be positive  at large $l$.

We can now review the situation in different cases. There is  no problem for dielectrics. At low temperatures only the low frequency photons with $\hbar \omega \sim k_{B}T$ are excited. Because  the imaginary part of the dielectric function $\varepsilon \prime  \prime (\omega) $ tends  to 0   when  $\omega \to 0$, these photons are well defined elementary excitations and their  thermodynamic functions behave in a usual way at $T \to 0$. 

The situation in conductors is more tricky. It was shown that for  metals with impurities, where the conductivity $\sigma(T) \to const $ as $T \to 0$, both 
 $\partial S_C/\partial l$ and $S_C$ tend to zero by the same law\cite{Hoye03,Bostrom04,Bezzera04}. Thus there is no any peculiarities again.
 
For pure metals, where the anomalous skin effect theory must be used, the situation is quite interesting. As it was shown 
by Svetovoy and Esquivel \cite{Svetovoy06}, in this case $|\partial S_C/\partial l| \propto T$, while $|S_C| \propto T^{2/3}$ (see next section). Thus $|S_C| \gg 
|\partial S_C/\partial l|$. A calculation gives that $S_C < 0$ and does not depend on $l$, as must be the case. The leading $l$-dependent term is proportional to $T$ and positive.

We see that there are no violations of the thermodynamic conditions in the Lifshitz theory of Casimir-Lifshitz forces for conductive media, if the problem is formulated and solved in a correct way. 

For {\it disordered} glass-like media  Nernst's theorem {\it is not valid}. They
are not at an equilibrium state at low temperatures
due to a very long relaxation time and have a large finite entropy
at zero temperature. (See, for example, a recent paper \cite{glasses} and references therein.) There is no reason why the difference (\ref{SC1}) must in this case tend to zero at $T \to 0$. 
 Thus general considerations cannot say anything about  the properties of $S_C$ and I will not discuss this case anymore. Notice only that the specific heat difference (\ref{DC}) must tend to zero also for disordered bodies. However, this condition is satisfied in a trivial way in all known examples due to the factor $T$ in the left hand side of the equation.

\section{The surface  entropy of a metal in condition of anomalous skin effect\label{B}}

As I noted previously, the presence of  the $l$-independent negative term in the ''Casimir entropy'' of a metal in conditions of  anomalous skin effect implies that  the free surface of this metal gives the main contribution to the entropy. It is instructive to examine this mechanism more closely by mens of a direct calculation of the entropy for a metallic half-space.  For this problem, the most convenient method is, in my opinion, the method of calculation of the Casimir-Lifshitz contribution to the free energy, which was suggested by Barash and Ginzburg   in 1975 \cite{BG75}. This powerful method, unfortunately, did not attract the attention which it surely deserves. (See, however, an interesting example of its use in Ref.  \cite{Brevik94}.) The method is analogous to the one presented in Ref. \cite{DLP61}. Differently from the latter, however, it does not demand the calculation of the full Green's function of electromagnetic field, but it only requires a dispersion relation for electromagnetic waves.  The dispersion relation can be presented in the form
 \begin{equation}
D(q,\omega)=0.
\label{disp}
\end{equation}
Then the Ginzburg-Barash equation \cite{BG75} for the contribution of the Van der Waals interactions to the free energy  can be written as:
\begin{eqnarray}
 F(T)&=&T\sum_{n=0}^{\infty}\,'\int d^2q\log D(q,i\zeta_n)\nonumber \\
&=&E(0)+\frac{1}{2\pi i}{\rm P}\int^{\infty}_{-\infty}\frac{d\omega}{e^{ \omega/T}-1}\int d^2q\log D(q,\omega)\nonumber \\
&=&E(0)+\frac{ T}{2\pi i}{\rm P}\int^{\infty}_ {-\infty}\frac{d\xi}{e^{ \xi}-1}\int d^2q\log D(q,T\xi),
\end{eqnarray}
where $q$ is the wave vector in the plane of the surface and $\zeta_n=2\pi Tn$.
Sometimes it is convenient to use the energy $E$ instead of the free energy. Taking into account that $E=F-T(\partial F/\partial T)$,
 one gets  the following contribution to the energy:
\begin{equation}
 E(T)=E(0)-\frac{ T^2}{2\pi i}\int^{\infty}_{-\infty}\frac{\xi d\xi}{e^{ \xi}-1}\int d^2q \frac{D'(q,T\xi)}{D(q,T\xi)}\; ,
\end{equation}
where 
\begin{equation}
D'= (\partial D/\partial \omega)\; .
\end{equation}

In the situation of the anomalous skin effect the dispersion relation must be expressed in terms of the surface impedance $Z(q,\omega)$: 
\begin{equation}
D(q,\omega) \propto \sqrt{\omega^2-q^2/c^2}Z(q,\omega)-i\omega/c=0 \; .\label{disp1}
\end{equation}
The impedance for waves of $s$-polarization has the form
\begin{equation}
Z(q,\omega) =(i\omega/qc)F(\omega \omega_p^2/q^3c^2v_F)\label{Z}
\end{equation}
where $\omega_p$ is the plasma frequency  and $v_F$ is the Fermi velocity of electrons in the metal. The function $F$ has been calculated
in \cite{Svetovoy05}. However, the goal of this section is to calculate only the temperature dependence of the $l$-independent term in $S_C$. Then we need not an explicit expression for $F$.

Notice first of all that in the calculation of the low temperature  behavior of the entropy the  values $\omega \sim T/\hbar \to 0$ are always important. Further,  it is obvious  from (\ref{disp1}) and (\ref{Z}) that the values $q \propto \omega^{1/3} \gg \omega $ are also important. Then we can neglect $\omega$ in comparison with $qc$ and $D \approx -(\omega/c)F(\omega \omega_p^2/q^3c^2v_F)$. The factor $(\omega/c)$ can be omitted and, finally, we find the estimate
\begin{equation}
 E(T)-E(0) \propto  T^2\int^{\infty}_{-\infty}\frac{\xi d\xi}{e^{ \xi}-1}\int \frac{d^2q}{q^3} \Phi \left (\frac{q}{(T\xi)^{1/3}} \right )
\propto T^{5/3}\; .\label{DEf}
\end{equation}
This estimate assumes that the integral converges. This can be easily checked. From (\ref{DEf}) we  finally find
that
\begin{equation}
S  \propto T^{2/3}\; .\label{Sf}
\end{equation}
It is worth noticing that we succeeded to calculate the leading term in the full entropy of the body only because of its singular behavior with respect to $T$. The singularity is produced by the contribution of small $q \propto T^{1/3}$ and the corresponding integrals converge. 
Notice also that the temperature dependence (\ref{Sf}) corresponds to the contribution to the entropy of surface waves with  dispersion law $\omega=Aq^{3}$. Of course,  under the considered conditions there are no undamped surface waves propagating along the surface. However, the evanescent waves give in this case the same type of contribution.
  
\section{Entropy of a helium film on a solid surface\label{C}}

In the previous section we have shown that in some conditions the temperature dependence of the entropy for a finite gap
between bodies can be different from one of a free surface. This result might look strange. However, in this section we will show  that the same situation takes place in a quite simple system - a superfluid film on a surface of a solid body, where the main contribution to the entropy is due to  quantized surface waves, so called ripplons.\cite{Atkins53} 

The surface waves on a surface of superfluid liquid are well-defined elementary excitations and their contribution to the free energy can be calculated in the usual way from the knowledge of the excitations energy spectrum $\omega=\omega(q)$:
 \begin{eqnarray}
 F &=&T\int \log \left ( 1-e^{-\omega(q)/T} \right)\frac{qdq}{2\pi} \nonumber \\
&=& - \frac{1}{4\pi} \int \frac{q^2d\omega}{e^{\omega/T}-1} \;.\label{Fs}
\end{eqnarray}
Neglecting  the effects of the gravity and of the Van der Waals interaction with the substrate, the waves on the surface of a liquid with thickness $l$ has the following dispersion law (see Problem 1 in \S 62, of Rev. \cite{LLFM}):
  \begin{equation}
\omega^2(q)=\frac{\alpha q^3}{\rho}\tanh(ql) \; ,
\end{equation}
where $\alpha$ is the surface tension of the liquid at $T=0$ and $\rho $ is its density. When $\l \to \infty $ we obtain the dispersion law for the free surface of a liquid half-space:
  \begin{equation}
\omega^2(q)=(\alpha q^3)/\rho \; .\label{disp1}
\end{equation}
Substitution (\ref{disp1}) into (\ref{Fs}) and integration give (in usual units) \cite{Atkins53}:
  \begin{equation}
F=-\Gamma(7/3)\zeta(7/3)\frac{(k_BT)^{7/3}\rho^{2/3}}{4\pi\hbar^{4/3}\alpha^{2/3}} \; .\label{F1}
\end{equation}
Correspondingly the entropy behaves as $S \propto T^{4/3}$. The values $q \sim (\rho/\alpha)^{1/3}(k_BT/\hbar)^{2/3} $ give the main contribution to the entropy and the result (\ref{F1}) is valid under the condition $k_BT \gg \hbar\alpha^{1/2}/(l^{3/2}\rho^{1/2})$.

In the opposite limit $ql \ll 1$ the dispersion law is:
  \begin{equation}
\omega^2(q)=(\alpha lq^4)/\rho \; 
\end{equation}
and
\begin{equation}
F=-\frac{\pi}{24}\frac{(k_BT)^{2}\rho^{1/2}}{\hbar\alpha^{1/2}l^{1/2}} \; .\label{F2}
\end{equation}
Correspondingly $S \propto T$. 
Again we find that the temperature dependence of the entropy is different for a free surface and a film of finite thickness. In the hydrodynamic example of surface waves the situation is opposite to the one of the Casimir-Lifshitz forces in section 2: here the film of finite thickness has larger entropy.

\section*{Acknowledgments}
I thank V.~Svetovoy for clarification of important points of previous papers and Yu.~Barash, F.~Dalfovo, C.~Henkel, F.~Intravaia and E.~Taylor for discussions. I acknowledge financial support by the European Science Foundation
(ESF) within the activity ''New Trends and Applications of the Casimir
Effect'' (www.casimir-network.com).

\end{document}